\documentclass[showpacs,twocolumn,aps,amssymb,preprintnumbers,
superscriptaddress,prc,floatfix]{revtex4}
\usepackage{epsfig,dcolumn}
\usepackage{graphicx,bm}
\usepackage{bm,color}
\newcommand{\be}{\begin{equation}}
\newcommand{\ee}{\end{equation}}
\newcommand{\ba}{\begin{eqnarray*}}
\newcommand{\ea}{\end{eqnarray*}}

\begin{document}
\title{A new effective interaction for 0$\bm{\hbar\omega}$ shell-model
       calculations in the $\bm{sd-pf}$ valence space}

\author{F. Nowacki}
\affiliation{IPHC, IN2P3-CNRS et Universit\'e Louis Pasteur, 
F-67037 Strasbourg, France}

\author{A. Poves}
\affiliation{Departamento de F\'isica Te\'orica e IFT-UAM/CSIC, 
Universidad Aut\'onoma de Madrid,  E-28049 Madrid, Spain}

\begin{abstract}

  The neutron-rich isotopes with Z$\le$20, in particular those with
  neutron numbers around N=28, have been at the focus of a lot
  experimental and theoretical scrutiny during the last few years.
  Shell-model calculations using the effective interaction SDPF-NR,
  were able to predict or to explain most of the properties featured
  by these nuclei. Prominent among them is the disappearance of the
  N=28 shell closure for Z$\le$16. We have incorporated 
  \textcolor{black}{into} SDPF-NR
  some modifications, either on purely theoretical grounds or guided
  by new experimental information. The proposed interaction SDPF-U
  offers enhanced reliability with respect to the earlier version.

\end{abstract}


\date{\today}
\maketitle

\section{Introduction}
\label{sec:intro}

In this article we describe the reshaping of the effective
interaction SDPF-NR, frequently used for 0$\hbar \omega$ shell model
calculations in the $sd$-$pf$ valence space. This interaction
originates in the work of Retamosa {\it et al.}  \cite{Ret97,Cau98}
and its final form was fixed in \cite{Num01}. More details about it
and some applications can be found in \cite{Cau05,Cau04}, and in
many of the experimental articles cited below.  From the very
beginning, the interest in the region was fostered by the experimental
indications of a weakening (or even a complete disappearance) of the
N=28 closure \cite{Sor93,Sor95,Gre04} and by the predictions of
several mean-field calculations in the same sense
\cite{Rei99,Lal99,Per00,Ray02}.  How can a shell closure vanish? Its
persistence depends on the balance of two opposite tendencies.
\textcolor{black}{On the one hand}, magic numbers are associated 
\textcolor{black}{with} energy gaps in the spherical mean field 
\textcolor{black}{and} this means that to promote particles above a 
closed shell costs energy. \textcolor{black}{On the other hand}, this 
energy can be partly recovered, because
closed shells have no correlation energy, while open-shell
configurations of neutrons and protons have a lot. Several  examples
of this phenomenon exist in stable nuclei in the form of coexisting
spherical, deformed and superdeformed states in a very narrow energy range.
A fully worked out case study in doubly magic $^{40}$Ca can be found in 
\cite{cmnp:07}. At the neutron-rich edge, the structure of the spherical 
mean field may be at variance with the usual one at the stability line. 
The reason is that,
at the stability line, the T=0 channel of the nucleon-nucleon
interaction has a stronger weight relative to the T=1 channel than it
has when the neutron excess is very large. When the gaps get reduced,
open-shell configurations, usually two-neutron excitations across the
neutron closure, take advantage of the availability of open-shell
protons to build highly correlated states that are more bound than the
closed-shell configuration.  Then the shell closure is said to have
vanished. $^{32}$Mg is a classical reference case for this kind of
physical behavior.
 
  The interaction SDPF-NR predicts a clear breaking of the N=28
magicity already at $^{44}$S.  The closed-shell configuration
$(0f_{7/2})^8$ represents only 24\% in $^{44}$S, 28\% in $^{42}$Si, and
just 3\% in $^{40}$Mg \cite{Cau04}. Its description of all the sulfur
isotopes is excellent \cite{Soh02} and its prediction of an isomer
0$^+$ state just above the first 2$^+$ has been verified experimentally
\cite{Gre05}. In addition, according to the calculations, the ground 
state of $^{43}$S
should be $\frac{3}{2}^-$ instead of the expected $\frac{7}{2}^-$,
suggesting that this nucleus is deformed. This has been verified
experimentally as well \cite{Sar00,Ibb99}. Many other properties are
also in agreement with the experimental data \cite{Dom03,Stu06}.

 The main reason that has prompted us to improve the existing effective 
interaction has been the vigorous renewal of interest in the region due 
to the access to many new neutron rich species,  in particular silicon 
isotopes, and indeed, because of the conflicting experimental 
\textcolor{black}{evidence} on the nature of $^{42}$Si coming from MSU and 
GANIL \cite{Fri05,Gad05,Gre05,Fri06,Jur06,Gau06,Gad06,Cam06,Bas07,Cam07}. 
Some of these experimental data will be analyzed in the final part of the
article. As more and more experimental results are accumulated, the need 
for a modified version of the SDPF-NR interaction becomes more compelling.

 As advanced in the \textcolor{black}{title}, our valence space
comprises the $sd$ shell for the valence protons, covering from Z=8 to
Z=20 and the  $sd$ and $pf$ shell for the neutrons, that is from N=8 to
N=40. Notice however that below N=20 these calculations reduce to the 
pure $sd$-shell ones, and beyond N=Z=20 to pure $pf$-shell ones.
The lightest nucleus in the valence space, that plays the role of
inert core, is therefore $^{16}$O.  The heaviest one is $^{60}$Ca. The
effective interaction is composed of three blocks; $sd$-$sd$,
 $sd$-$pf$ (proton-neutron), and $pf$-$pf$
(neutron-neutron). All the shell-model calculations are unrestricted
in the full $sd$ shell for the protons and the full $pf$ shell for the
neutrons.

\section{The reference states}
\label{sec2}

   In order to improve the monopole part of the SDPF-NR interaction,
we rely on the new experimental information coming from nuclei that have one
particle or one hole on the top of neutron and proton closures, as
well as on the values of the quasi-particle gaps across these
closures. Indeed, the correlations are fully taken into account. This
means that in order to find the right monopole changes several
iterations are usually needed. The results for the reference states,
using the new interaction SDPF-U, are compared in Table \ref{tab:t1}
with the experimental data (taken from the NNDC database \cite{nndc}
when no explicit reference is given below) and with the predictions of its
precursor interaction, SDPF-NR. The neutron gap in $^{48}$Ca is 4.74~MeV, 
compared with the experimental value,  4.78~MeV.

\begin{table}
\caption{Predictions of the SDPF-U interaction for the excitation energies 
of the reference states, compared with the experimental data and with the 
predictions of the SDPF-NR interaction (in MeV).}
\label{tab:t1}
    \begin{tabular*}{\linewidth}{@{\extracolsep{\fill}}lclll}
\hline \noalign{\smallskip}
 & J$^{\pi}$ & EXP. & SDPF-U & SDPF-NR \\  
\noalign{\smallskip} \hline \noalign{\smallskip}
 $^{35}$Si & $\frac{7}{2}^-$ & 0.0   & 0.0  & 0.0    \\ 
           & $\frac{3}{2}^-$ & 0.91  & 0.93 & 0.95   \\ 
           & $\frac{1}{2}^-$ & (2.0) & 2.14 & 3.79   \\ 
           & $\frac{5}{2}^-$ & ---   & 4.16 & 4.76   \\ [5pt]

 $^{35}$P  & $\frac{1}{2}^+$ & 0.0  & 0.0  & 0.0      \\ 
           & $\frac{3}{2}^+$ & 2.39 & 2.50 & 2.65     \\ 
           & $\frac{5}{2}^+$ & 3.86 & 3.91 & 4.27     \\ [5pt]

 $^{37}$S & $\frac{7}{2}^-$ & 0.0  & 0.0  & 0.0  \\ 
          & $\frac{3}{2}^-$ & 0.65 & 0.62 & 0.62 \\ 
          & $\frac{5}{2}^-$ & 2.51 & 2.44 & 2.44 \\ 
          & $\frac{1}{2}^-$ & 2.64 & 2.61 & 2.61 \\ [5pt]

 $^{39}$Ar & $\frac{7}{2}^-$ & 0.0  & 0.0  & 0.0     \\ 
           & $\frac{3}{2}^-$ & 1.27 & 1.54 & 1.37    \\ 
           & $\frac{5}{2}^-$ & 2.09 & 2.13 & 2.12    \\ 
           & $\frac{1}{2}^-$ & ---  & 3.08 & 3.02    \\ [5pt]

  $^{41}$Ca & $\frac{7}{2}^-$ & 0.0  & 0.0  & 0.0  \\ 
            & $\frac{3}{2}^-$ & 2.50 & 2.50 & 1.98 \\ 
            & $\frac{1}{2}^-$ & 4.16 & 4.20 & 3.97 \\ 
            & $\frac{5}{2}^-$ & 6.98 & 6.99 & 6.49 \\ [5pt]

 $^{47}$Ar & $\frac{3}{2}^-$ & 0.0  & 0.0  & 0.0  \\ 
           & $\frac{1}{2}^-$ & 1.13 & 1.14 & 1.28 \\ 
           & $\frac{5}{2}^-$ & ---  & 1.28 & 1.12 \\ 
           & $\frac{7}{2}^-$ & 1.74 & 1.58 & 1.37 \\ [5pt]

 $^{47}$K & $\frac{1}{2}^+$  & 0.0  & 0.0  & 0.0     \\ 
           & $\frac{3}{2}^+$ & 0.36 & 0.32 & 0.31    \\ 
           & $\frac{5}{2}^+$ & 3.32 & 3.00 & 2.97    \\ [5pt]

 $^{49}$K & $\frac{3}{2}^+$ & (0.0)  & 0.0  & 0.074  \\ 
          & $\frac{1}{2}^+$   & --- & 0.08 & 0.0    \\ 
          & $\frac{5}{2}^+$   & --- & 0.70 & 0.73   \\ [5pt]

 $^{49}$Ca & $\frac{3}{2}^-$ & 0.0  & 0.0  & 0.0    \\ 
           & $\frac{1}{2}^-$ & 2.02 & 1.96 & 1.70   \\ 
           & $\frac{7}{2}^-$ & 3.35 & 3.14 & 2.46   \\ 
           & $\frac{5}{2}^-$ & 3.59 & 3.88 & 3.20    \\ 
           & $\frac{5}{2}^-$ & 3.99 & 4.03 & 3.83   \\ [5pt]
\noalign{\smallskip}\hline
\end{tabular*}
\end{table}

The nucleus $^{35}$Si plays a pivotal role in fixing the evolution of
the effective single-particle energies in this space. The reason is
twofold; it is very single-particle-like, because of the doubly-magic
character of $^{34}$Si, and its neutron single-particle states define
the location of the $pf$ orbits at the neutron-rich side. That is why
the changes made in the initial interaction of Retamosa {\it et al.}
\cite{Ret97} to comply with the experimental finding of the low
excitation energy of the $\frac{3}{2}^-$ state of $^{35}$Si
\cite{Num01} had \textcolor{black}{such} important consequences in some 
cases.  The modifications leading to the interaction SDPF-NR  
left the calculated centrois of the $\frac{1}{2}^-$ and $\frac{5}{2}^-$ states 
of $^{35}$Si at their initial \textcolor{black}{positions, {\it i.e.}
similar to the ones} they have in $^{41}$Ca. Besides that, in order to 
place the $\frac{3}{2}^-$ state of $^{35}$Si 
at its experimental excitation energy, only the
0$d_{3/2}$-1$p_{3/2}$ centroid was modified. As a consequence,  SDPF-NR 
produces a $\frac{1}{2}^+$ ground state in  $^{49}$K,  
instead of the $\frac{3}{2}^+$ favored by the experiments \cite{nndc},
and does not reproduce the assignment of
J=0$^-$ to the ground state of $^{50}$K proposed in \cite{Bau98}.

  Two more pieces of experimental information are considered as well;
the recent study of the single-particle states in $^{47}$Ar \cite{Gau06} 
that has shown a certain reduction of the
spin-orbit splitting of the $p$ orbits compared with the situation in
$^{49}$Ca, and  the spectrum of  $^{37}$S \cite{End88}.
 Finally, the cross-shell monopoles were also constrained by requiring
exact agreement with the pf-shell centroids in $^{41}$Ca extracted
from the $^{40}$Ca(d,p) reaction \cite{Uoz:94}. 

 With all these boundary conditions, the monopole part of the
interaction would be uniquely defined, provided we knew the full
$^{35}$Si single-particle spectrum, \textcolor{black}{which} we do
not.  Among the possible choices, the one that fits
\textcolor{black}{most} naturally with the available experimental
indications \cite{Gel:07} is that of a compressed spectrum with
excitation energies 1~MeV, 2~MeV and 4~MeV for the $\frac{3}{2}^-$,
$\frac{1}{2}^-$, and $\frac{5}{2}^-$ states. This time, we produce the
compression by changing equally the 0$d_{5/2}$-r$_3$, 0$d_{3/2}$-r$_3$
and 1$s_{1/2}$-r$_3$ centroids (r$_p$ is a shorthand notation for all
the orbits in the major oscillator shell of energy $\hbar
\omega(p+\frac{3}{2})$ except the one with the largest $j\!=\!p+1/2$).
In order to obtain reasonable binding-energy differences, we have
adjusted the global French-Bansal monopole parameter a=$
\frac{1}{4}(3\overline{\rm V}^1$+$\overline{\rm V}^0$) of the
$sd$-$pf$ interaction to the one-neutron separation energy of
$^{41}$Ca (8.35~MeV) and the total centroid of the neutron-neutron
$pf$-$pf$ interaction to the $^{52}$Ca -- $^{40}$Ca mass
difference. This translates into  a modification of all the T=0 and T=1
$sd$-$pf$ centroids of --0.045~MeV and a modification of all the T=1
$pf$-$pf$ centroids of --0.250~MeV.  If the experimental data on
separation energies would demand it, extra terms quadratic in the
number of $sd$ valence protons, n$_{\pi}$=(Z-8), and $pf$ valence
neutrons, n$_{\nu}$=(N-20), could be added to the interaction without
changing its spectroscopic properties

\begin{figure}
\begin{center}
\includegraphics[width=1.0\columnwidth,angle=0]{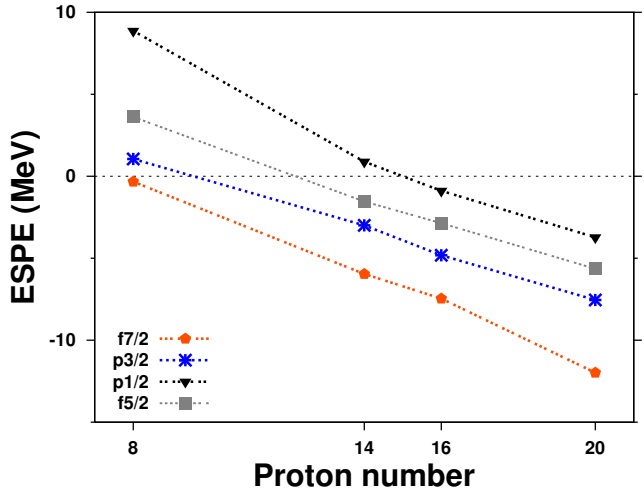}
\end{center}
\caption{(Color online) Effective neutron single-particle energies at 
N=28 from Z=8 to Z=20 \label{fig:n28}.}
\end{figure}

\begin{figure}
\begin{center}
\includegraphics[width=1.0\columnwidth,angle=0]{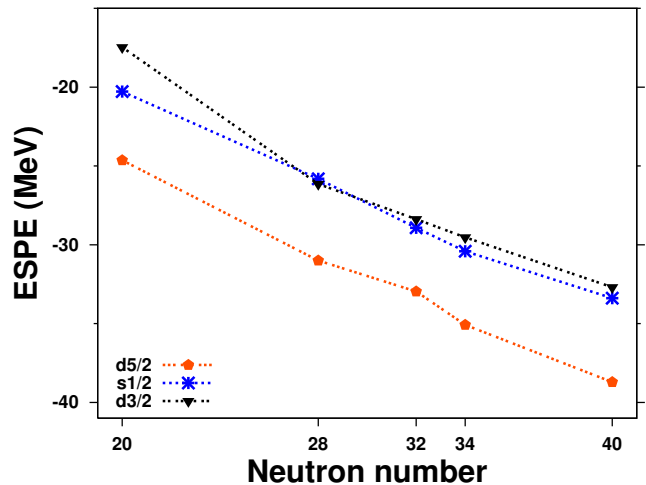}
\end{center}
\caption{(Color online) Effective proton single-particle energies in the 
calcium isotopes from N=20 to N=40 \label{fig:z20}.}
\end{figure}

 The spherical mean field that is produced by the new interaction is
better visualized by the evolution of the effective single-particle
energies (ESPE) with varying proton and neutron numbers.  These are
presented in Figs.~\ref{fig:n28} and \ref{fig:z20},
where we plot the $pf$-shell neutron ESPE's at N=28 and Z=8-20, and the
$sd$-shell proton ESPE's at Z=20 and N=20-40, respectively.  In
Fig.~\ref{fig:n28}, we can follow the reduction of the N=28 neutron gap
as we approach the neutron-rich edge. This reduction, combined with
the degeneracy of the 0d$_{3/2}$ and 1s$_{1/2}$ proton orbits at N=28,
shown in Fig.~\ref{fig:z20}, is at the origin of the vanishing of the
N=28 magicity for Z=16. The reason is that when the 0d$_{3/2}$ and
1s$_{1/2}$ are degenerate, they form a pseudo-SU3 doublet that
enhances the quadrupole correlations of the configurations with 
open-shell neutrons.  Notice also that beyond N=28 the orbits cross 
again recovering their standard ordering. The monopole part of the 
SDPF-U interaction (without the two global modifications discussed above)
is collected in Table~\ref{tab:mono} at the end of the paper, and 
compared with the monopole part of the SDPF-NR interaction.

\section{Pairing and the Silicon chain}
\label{sec3}
The excitation energies of the first
2$^+$ states of the N=22 isotones (two neutrons in the $pf$-shell)
are predicted too high by the SDPF-NR interaction. The 
deviations increase as protons are removed from $^{42}$Ca till $^{36}$Si.
The 2$^+$ states of  $^{38}$Si and $^{40}$Si are also too high by about
400~keV (this fact as been also discussed recently in \cite{Cam06}).
However, the same interaction gives the right excitation
energy for the very neutron-rich  sulfur isotopes.  What is the origin of this
discrepancy? In our opinion\textcolor{black}{, it} is a subtle point of 
effective interaction
theory. The $pf$-shell effective interaction contained in SDPF-NR is
renormalized in order to compensate for the absence of 2p-2h
excitations from the core. This renormalization, that is predominantly
pairing-like, and that seems to have the right size for Z$>$14, depends on
the value of the $^{40}$Ca gap.
 However, when we deal with the silicon isotopes, the relevant
gap for a perturbative estimation of the pairing renormalization is
the splitting of the 0$d_{5/2}$ and 0$f_{7/2}$ proton orbits in $^{34}$Si, 
which may double the former one. In addition,
the off-diagonal pairing matrix element $\langle 0d_{3/2}^2 (J=0)|V|
0f_{7/2}^2 (J=0) \rangle$ is larger than $\langle 0d_{5/2}^2 (J=0)|V|
0f_{7/2}^2 (J=0) \rangle$. Using the values of the gaps and these
matrix elements, we have estimated the excess of pairing renormalization 
for Z$\le$14 \textcolor{black}{to be} $\approx$250~keV. Therefore, we have
removed from SDPF-NR a schematic pairing Hamiltonian in the $pf$-shell
with G=-0.09~MeV, which, for the $\langle 0f_{7/2}^2 \mbox{\rm
(J=0,T=1)}|V|0f_{7/2}^2 \mbox{\rm (J=0, T=1)}\rangle$ matrix element
represents roughly 250~keV.  With this change, the 2$^+$ excitation
energies of the silicon isotopes agree nicely with the
experiment. Therefore, we have been led, and for good reasons, to have
two interactions one for Z$>$14 and another  for Z$\le$14.  The
files with the two versions of the SDPF-U interaction will be available
with the interaction package attached to the open version of the code
{\sf ANTOINE}, that can be downloaded from the URL
\url{http://sbgat194.in2p3.fr/~theory/antoine/main.html}.

\section{The Z=14 gap; deformation in $^{42}$Si}
\label{sec4}

\begin{table}
\caption{Properties of  $^{42}$Si for different effective interactions. 
Energies in MeV, B(E2)'s in e$^2$~fm$^4$ and Q's in e~fm$^2$.} 
\label{tab:t2}
    \begin{tabular*}{\linewidth}{@{\extracolsep{\fill}}lccc}
\hline \noalign{\smallskip}
 &  SDPF-NR
 &  SDPF-NR$^{(a)}$
 &  SDPF-U  \\   
 \noalign{\smallskip}\hline\noalign{\smallskip}
Z=14 gap                         & 6.08 & 6.22 & 5.36 \\ 
n$_{\pi}$(0d$_{\frac{5}{2}}$)    & 5.42 & 5.43 & 4.83 \\ 
N=28 gap                         & 3.50 & 3.20 & 2.75 \\ 
n$_{\nu}$(0f$_{\frac{7}{2}}$)    & 6.16 & 6.55 & 6.08 \\ 
\hline \noalign{\smallskip}
E$^*$(2$^+$)                & 1.49 & 1.25 & 0.82 \\
Q$_s$                  &  15   & 16    & 20   \\
BE2$\downarrow$        &  53  & 48    & 86  \\
Q$_0$ (from Q$_s$)     & -53  & -56  & -70   \\
Q$_0$ (from BE2)       & -52 &  -50  & -67   \\
\noalign{\smallskip}
\hline
\end{tabular*}
    \begin{tabular*}{\linewidth}{@{\extracolsep{\fill}}l}
$^{(a)}$ including the pairing correction discussed in the text. \\
\end{tabular*}
\end{table}

 As we mentioned in the introduction, there has been recently an
interesting debate on the nature of $^{42}$Si, following the claim of
its magic character \textcolor{black}{in Ref.}~\cite{Fri05}. In a subsequent 
paper by the same authors \cite{Fri06}, the claim was essentially suspended as
conflicting \textcolor{black}{evidence} appeared. A gamma ray of very low energy
observed at GANIL was attributed to the decay of the first excited
2$^+$ state. Such a low-lying 2$^+$ state is totally incompatible with
a magic structure in this mass region \cite{Gre05,Bas07}.  The
interaction SDPF-NR, already indicated the vanishing of the N=28 shell
closure and the development of oblate collectivity in $^{42}$Si. Its
predictions for $^{42}$Si are gathered in the first entry of
Table~\ref{tab:t2}. Even if the excitation energy of the 2$^+$ state
is not low enough, the quadrupole transitions are large and consistent
with the predicted (oblate) spectroscopic quadrupole moment. Besides,
the occupancy\textcolor{black}{, 6.16,} of the 0f$_{\frac{7}{2}}$ neutron 
orbit is very low, corresponding to a \textcolor{black}{closed-shell 
percentage} of 28\%. 
This evidence was somehow masked by the too strong $pf$-shell pairing that
we have discussed in the previous section, as can be seen clearly in
the column of \textcolor{black}{Table~\ref{tab:t2}} labeled \mbox{SDPF-NR$^{(a)}$} 
(the SDPF-NR interaction including the pairing modification mentioned 
before).  Even if the closed-shell probability increases, the quadrupole 
collectivity \textcolor{black}{remains} and the 2$^+$ excitation energy is 
now much lower.  The other difference between the SDPF-NR interaction 
and the new SDPF-U \textcolor{black}{interaction} is
related to the evolution of the Z=14 proton gap from N=20 to N=28.  In
fact, with the interaction SDPF-NR, the centroid of the (proton)
0d$_{\frac{5}{2}}$ spectroscopic strength in $^{47}$K is too high in
energy compared with the experimental value extracted in
~\cite{Kra01}. The effect of reducing this gap is to lower the
2$^+$ excitation energy and to increase the quadrupole
collectivity. Eventually, the SDPF-U interaction makes $^{42}$Si a 
well-deformed oblate rotor with $\beta$$\sim$0.4.

In  Table  \ref{tab:t3}  we  have  gathered  the  basic  spectroscopic
predictions  of the new  effective interaction  for some  very neutron
rich silicon isotopes \textcolor{black}{and compared them} with the available 
experimental data. The quadrupole properties are computed  with the effective
charges q$_{\nu}$=0.35, q$_{\nu}$=1.35 that give the best reproduction
of  the quadrupole  properties of  the $sd$\textcolor{black}{-shell}  
nuclei  \cite{Bro88}. The
agreement with the available data  is very satisfactory.  The two most
neutron-rich  isotopes show very  distinct and peculiar  behaviors. In
$^{42}$Si, the yrast  sequence 0$^+$, 2$^+$, 4$^+$ does  not follow the
rotational J(J+1) law. \textcolor{black}{However,} its quadrupole 
properties (B(E2)'s and
spectroscopic quadrupole moments) are consistent with the existence of
an   intrinsically-deformed  oblate   structure.  In   addition,  the
calculation produces a low-lying excited 0$^+$ state at 1.08 MeV. This
state, the  second-excited 2$^+$  state, and the second-excited 4$^+$
state are connected by E2 transitions of the of the same strength 
\textcolor{black}{as} those of the  yrast band, but this time,  the 
spectroscopic quadrupole moments  have similar absolute  values but  
opposite sign. \textcolor{black}{This is a} blatant
case of coexistence  of two deformed intrinsic states,  one oblate and
another prolate.  In $^{40}$Si, the B(E2) values of the yrast band are
smaller than  the corresponding ones  in $^{42}$Si, while  still being
substantial. The calculation produces an excited 0$^+$ state that lies
higher (1.82 MeV)  and a very low-lying second-excited 2$^+$ state at
1.25 MeV.  This 2$^+$ is  connected by a  strong E2 transition  to the
first 3$^+$  state at  1.8 MeV, forming  a sort of  incipient $\gamma$
band.

\begin{table}
\caption{Basic spectroscopy of the silicon isotopes. Energies are in MeV, 
B(E2)'s in e$^2$~fm$^4$ and Q's in e~fm$^2$.} 
\label{tab:t3}
    \begin{tabular*}{\linewidth}{@{\extracolsep{\fill}}lccccc}
\hline \noalign{\smallskip}
 &  A=36
 &  A=38
 &  A=40
 &  A=42
 &  A=44 \\  
 \noalign{\smallskip}\hline\noalign{\smallskip}
E$^*$(2$^+_1$) th. & 1.36 & 0.96 & 0.80  & 0.82 & 0.87 \\
E$^*$(2$^+_1$) exp. & 1.40 & 1.08 & 0.99  & 0.77 &  \\
BE2($2^+_1 \rightarrow 0^+_1$) th.    &   34 &  39 &   54 &  86 & 44  \\
BE2($2^+_1 \rightarrow 0^+_1$) exp.    &   39(12) &  38(14) &    &   &   \\
Q$_s$(2$^+_1$) th.               &   +3 &  -9 &   -9 &  +20 & +4   \\
E$^*$(4$^+_1$) th. & 2.74 & 1.83 & 1.99  & 1.79 & 2.53 \\
S$_{2n}$ th.  & 9.82  & 8.33   & 7.04 &  5.61  & 3.47  \\
\noalign{\smallskip}
\hline
\end{tabular*}
\end{table}

\section{The very heavy magnesiums}
\label{sec6}

 The spectroscopy of the magnesium isotopes has unveiled many cases
of unexpected physical behavior. The N=Z member of the chain,
$^{24}$Mg, is the paradigm of a well-deformed nucleus that can be
described in the laboratory frame by means of Elliott's SU3 model.  
\textcolor{black}{ At N=20, $^{32}$Mg is not a semi-magic nucleus but, 
on the contrary, a well-deformed prolate rotor}. It provides one of the first
examples of the vanishing of a magic closure by the action of deformed
intruders \cite{Gui84,Mot95,Pov87}.  As can be seen in table
\ref{tab:t4}, the scarce experimental information on $^{34}$Mg
\cite{Yon01,Iwa01,Chu05,Ele06} and $^{36}$Mg \cite{Gade07} also point
to well-deformed cases. We produce a $^{34}$Mg  less collective than the
experiment, surely due to the absence of neutron excitations from the
$sd$-shell to the $pf$-shell in our valence space.
 In $^{36}$Mg, the 0$\hbar \omega$ calculation already
overshoots the lowering of the 2$^+$ state \textcolor{black}{and}, 
therefore, it does not
seem necessary to resort to a large intruder mixing to explain it.
When more neutrons are added, $^{38}$Mg, and, prominently $^{40}$Mg,
are extremely deformed prolate rotors with very low 2$^+$ states.
$^{40}$Mg is at the edge of the neutron drip line, and, according to
our calculations, it has nearly two neutrons in the 1p$_{\frac{3}{2}}$
orbit, a situation that may favor the formation of a neutron halo;
$^{40}$Mg could then be the first well deformed nucleus adorned with a
neutron halo. In addition, the calculations suggest that in $^{40}$Mg,
the prolate yrast band may coexist with an oblate one based in an 
excited  0$^+$ state at 2.07 MeV, in parallel with the situation found 
in  $^{42}$Si. $^{38}$Mg is, according to our description, a triaxial
rotor in view of the presence of a $\gamma$ band based upon the second
 2$^+$ state at 1.07~MeV excitation energy. This band comprises the
first  3$^+$ and the second  4$^+$ states. The energies and the ratios of the
transitions between the two bands are consistent with $\gamma$=0.2.
Similar triaxial structures have been recently shown in the heaviest
Argon isotopes~\cite{bh08}.

\begin{table}
\caption{Basic spectroscopy of the Magnesium isotopes. Energies are in MeV, 
B(E2)'s in e$^2$~fm$^4$ and Q's in e~fm$^2$.} 
\label{tab:t4}
    \begin{tabular*}{\linewidth}{@{\extracolsep{\fill}}lcccc}
\hline \noalign{\smallskip}
 &  A=34
 &  A=36
 &  A=38
 &  A=40 \\  
 \noalign{\smallskip}\hline\noalign{\smallskip}
E$^*$(2$^+_1$) th. & 0.82 & 0.48 & 0.55  & 0.55   \\
E$^*$(2$^+_1$) exp. & 0.67 & 0.66 &  &   \\
BE2$\downarrow$ th.    &  65 &   88 &  101 &  98    \\
BE2$\downarrow$ exp.    &  110(20) &   &    &      \\
Q$_s$               &   -15 &  -19 &   -18 &  -20     \\
E$^*$(4$^+_1$)       & 1.99 & 1.39 & 1.60 & 1.77   \\
S$_{2n}$ th.  & 7.76 & 6.50   & 4.29 & 2.67    \\
\noalign{\smallskip}
\hline
\end{tabular*}
\end{table}

\section{The evolution of the $\frac{1}{2}^+$--$\frac{3}{2}^+$ 
splitting  in the K, Cl and P isotopes}
\label{sec5}

\begin{figure}
\begin{center}
\includegraphics[width=1.0\columnwidth,angle=0]{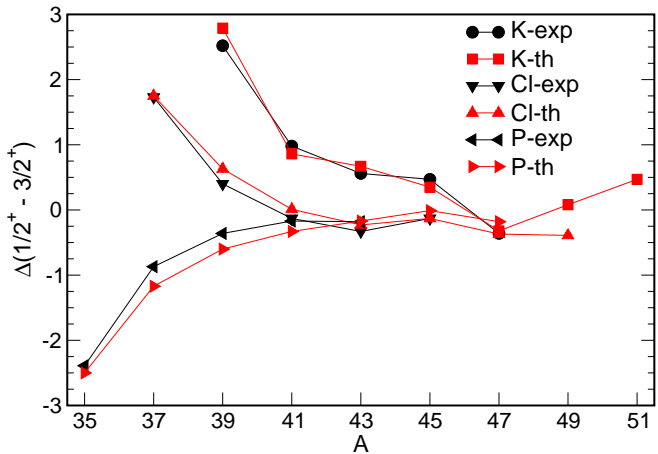}
\end{center}
\caption{(Color online) Energy splitting (in MeV) between the lowest
 $\frac{1}{2}^+$ and $\frac{3}{2}^+$ states in the neutron-rich isotopes
of potassium, chlorine, and phosphorus \label{fig:holes}.}
\end{figure}

  In Fig.~\ref{fig:holes}, we present the evolution of the
 energy splitting between the lowest $\frac{1}{2}^+$ and
 $\frac{3}{2}^+$ states in the neutron-rich isotopes of potassium,
 chlorine and phosphorus.  Notice in the first place the excellent
 agreement of the theoretical predictions with the available
 experimental data.  In the limit of pure single-particle behavior,
 the points should \textcolor{black}{lie on} straight lines joining the 
 neutron closures
 N=20, N=28 and N=32, corresponding to the ones in Fig.~\ref{fig:z20}
 for the potassium chain. This is clearly not the case, because of the
 coupling of the proton-single particle states with the neutron
 excitations. As we can gather from the figure, the splitting in the
 potassiums drops rapidly when two neutrons are added to the N=20
 closure, meaning that the states have not a single-particle character 
 anymore.  At N=28 the single-particle behavior is recovered and the 
 inversion of  the two states that takes place  is  dictated by the effective 
 single-particle energies.
 The crossing of the two orbits just reflects the fact
 that the neutron-proton interaction is more attractive between the
 orbits 0f$_{\frac{7}{2}}$ and 0d$_{\frac{3}{2}}$ than between the
 orbits 0f$_{\frac{7}{2}}$ and 1s$_{\frac{1}{2}}$. When neutrons fill
 the 1p$_{\frac{3}{2}}$ orbit, the opposite is true and the ground-state 
 spins  of $^{49}$K and $^{51}$K bounce back to $\frac{3}{2}^+$. The
 situation is less clear cut in the chlorines, because its three
 proton holes produce enhanced mixing. This can be seen also in 
  Fig.~\ref{fig:holes}; at N=20, $^{37}$Cl has the normal ordering of 
 spins that  corresponds to a ground-state proton configuration
 (1s$_{\frac{1}{2}}$)$^2$ (0d$_{\frac{3}{2}}$)$^1$ J= $\frac{3}{2}^+$
 with an excited $\frac{1}{2}^+$ state corresponding to the
 configuration (1s$_{\frac{1}{2}}$)$^1$ (0d$_{\frac{3}{2}}$)$^2$.
 Increasing the number of neutrons reduces the
 0d$_{\frac{3}{2}}$-1s$_{\frac{1}{2}}$ splitting and increases the
 mixing among different proton configurations.  Already at N=26 both
 states approach the configuration (1s$_{\frac{1}{2}}$)$^1$
 (0d$_{\frac{3}{2}}$)$^2$ and become nearly degenerate as the protons
 enter in a pseudo-SU3 regime. The heavy chlorines may be viewed as
 one proton coupled to the corresponding (A-1) sulfur isotope.  At
 N=26, $^{42}$S is deformed and at N=28 the neutron closure has
 vanished in $^{44}$S; this explains why, in the chlorine chain, there
 is no trace of the single-particle effects that are so visible in the
 potassium chain when we cross N=28.  {\it Mutatis mutandis}, these
 arguments hold for the phosphorus isotopes.

\begin{table}
\caption{Centro\"{\i}ds of the SDPF-U(*) and SDPF-NR interactions for A=18 (in MeV).
 The standard mass dependence 
(A/18)$^{1/3}$ is adopted in the calculations.
The single particle energies (in MeV) on a core of $^{16}$O are the following:
0d$_{5/2}$=--3.70;  1s$_{1/2}$=--2.92;   0d$_{3/2}$=1.90;
     0f$_{7/2}$=6.22;     1p$_{3/2}$=6.31;      0f$_{5/2}$=11.45;    1p$_{1/2}$=6.48.}
\label{tab:mono}
    \begin{tabular*}{\linewidth}{@{\extracolsep{\fill}}rrlrr}
\hline \noalign{\smallskip}
 j$_1$ & j$_2$ & T & SDPF-U & SDPF-NR \\  
\noalign{\smallskip} \hline \noalign{\smallskip}
    0d$_{5/2}$ &  0d$_{5/2}$ &     0  &  -3.19 &  -3.19 \\
    0d$_{5/2}$ &  0d$_{5/2}$ &     1  &  -0.53 &  -0.61 \\
    0d$_{5/2}$ &  1s$_{1/2}$ &     0  &  -3.12 &  -3.12 \\
    0d$_{5/2}$ &  1s$_{1/2}$ &     1  &   0.14 &   0.14 \\
    0d$_{5/2}$ &  0d$_{3/2}$ &     0  &  -3.74 &  -3.74 \\
    0d$_{5/2}$ &  0d$_{3/2}$ &     1  &  -0.33 &  -0.29 \\
   1s$_{1/2}$  &  1s$_{1/2}$ &     0  &  -3.51 &  -3.51 \\
   1s$_{1/2}$  &  1s$_{1/2}$ &     1  &  -2.15 &  -2.15 \\
   1s$_{1/2}$  &  0d$_{3/2}$ &     0  &  -3.01 &  -3.01 \\
   1s$_{1/2}$  &  0d$_{3/2}$ &     1  &   0.00 &   0.00 \\
    0d$_{3/2}$ &  0d$_{3/2}$ &     0  &  -2.67 &  -2.67 \\
    0d$_{3/2}$ &  0d$_{3/2}$ &     1  &  -0.31 &  -0.40 \\ 
\noalign{\smallskip}\hline \noalign{\smallskip}
    0d$_{5/2}$ &  0f$_{7/2}$ &     0  & -2.55  & -2.85 \\
    0d$_{5/2}$ &  0f$_{7/2}$ &     1  &  0.00  &  0.00 \\
    0d$_{5/2}$ &  1p$_{3/2}$ &     0  & -1.51  & -2.50 \\
    0d$_{5/2}$ &  1p$_{3/2}$ &     1  & -0.31  & -0.08 \\
    0d$_{5/2}$ &  0f$_{5/2}$ &     0  & -3.17  & -4.16 \\
    0d$_{5/2}$ &  0f$_{5/2}$ &     1  & -0.32  & -0.09 \\
    0d$_{5/2}$ &  1p$_{1/2}$ &     0  & -1.77  & -3.25 \\
    0d$_{5/2}$ &  1p$_{1/2}$ &     1  & -0.48  & -0.08 \\
   1s$_{1/2}$  &  0f$_{7/2}$ &     0  & -1.89  & -1.89 \\
   1s$_{1/2}$  &  0f$_{7/2}$ &     1  & -0.30  & -0.30 \\
   1s$_{1/2}$  &  1p$_{3/2}$ &     0  & -1.44  & -1.07 \\
   1s$_{1/2}$  &  1p$_{3/2}$ &     1  & -1.12  & -1.41 \\
   1s$_{1/2}$  &  0f$_{5/2}$ &     0  & -2.48  & -3.17 \\
   1s$_{1/2}$  &  0f$_{5/2}$ &     1  & -0.05  &  0.18 \\
   1s$_{1/2}$  &  1p$_{1/2}$ &     0  & -1.91  & -4.21 \\
   1s$_{1/2}$  &  1p$_{1/2}$ &     1  &  0.05  &  0.43 \\
    0d$_{3/2}$ &  0f$_{7/2}$ &     0  & -2.43  & -2.43 \\
    0d$_{3/2}$ &  0f$_{7/2}$ &     1  & -0.88  & -0.88 \\
    0d$_{3/2}$ &  1p$_{3/2}$ &     0  & -1.72  & -2.26 \\
    0d$_{3/2}$ &  1p$_{3/2}$ &     1  & -0.30  & -0.15 \\
    0d$_{3/2}$ &  0f$_{5/2}$ &     0  & -2.39  & -3.08 \\
    0d$_{3/2}$ &  0f$_{5/2}$ &     1  &  0.27  &  0.50 \\
    0d$_{3/2}$ &  1p$_{1/2}$ &     0  & -2.05  & -2.79 \\
    0d$_{3/2}$ &  1p$_{1/2}$ &     1  &  0.03  &  0.41 \\ 
\noalign{\smallskip}\hline \noalign{\smallskip}
    0f$_{7/2}$ &  0f$_{7/2}$ &     1  & -0.27  &  -0.37 \\
    0f$_{7/2}$ &  1p$_{3/2}$ &     1  &  0.15  &   0.12 \\
    0f$_{7/2}$ &  0f$_{5/2}$ &     1  &  0.03  &  -0.04 \\
    0f$_{7/2}$ &  1p$_{1/2}$ &     1  &  0.20  &   0.07 \\
   1p$_{3/2}$  &  1p$_{3/2}$ &     1  & -0.10  &  -0.81 \\
   1p$_{3/2}$  &  0f$_{5/2}$ &     1  &  0.36  &  -0.16 \\
   1p$_{3/2}$  &  1p$_{1/2}$ &     1  &  0.16  &  -0.62 \\
    0f$_{5/2}$ &  0f$_{5/2}$ &     1  &  0.12  &  -0.07 \\
    0f$_{5/2}$ &  1p$_{1/2}$ &     1  &  0.14  &  -0.05 \\
   1p$_{1/2}$  &  1p$_{1/2}$ &     1  &  0.25  &  -0.46 \\ 
\noalign{\smallskip}\hline
\end{tabular*}
(*) without the global monopole corrections (see text).
\end{table}

\section{Conclusions}
\label{sec7}

  In summary, we have produced a new effective interaction, SDPF-U, suitable
for unrestricted 0$\hbar \omega$ large-scale shell-model calculations in the
$sd-pf$ valence space, for nuclei with proton number ranging from Z=8
to Z=20 and neutron number from N=20 to N=40.  We have presented a first
series of theoretical predictions and have compared them with the 
experimental results.
Special attention has been devoted to the behavior of the neutron rich
silicon and magnesium isotopes and to the vanishing of the N=28 neutron
shell closure in  $^{42}$Si and $^{40}$Mg. We have also explored the evolution
of the  energy splitting between the lowest $\frac{1}{2}^+$ and
 $\frac{3}{2}^+$ states in the neutron-rich isotopes of potassium,
chlorine and phosphorus, finding an excellent agreement with the experimental
results.

{\bf Acknowledgments.} The authors express their warmest thanks to
                       Etienne Caurier for his help in different
                       aspects of this work. Partly supported  by a grant of
                       the Spanish Ministry of Education and Science
                       (FPA2007-66069), by the IN2P3(France)-CICyT(Spain)
                       collaboration agreements, by the Spanish
                       Consolider-Ingenio 2010 Program CPAN
                       (CSD2007-00042) and by the Comunidad de Madrid
                       (Spain), project HEPHACOS P-ESP-00346.


\begin{thebibliography}{99}


\bibitem{Ret97} J. Retamosa, E. Caurier, F. Nowacki, and A. Poves, 
              Phys. Rev. C \textbf{55}, 1266 (1997).

\bibitem{Cau98} E. Caurier, F. Nowacki,  A. Poves, and J. Retamosa, 
               Phys. Rev. C \textbf{58}, 2033 (1998).

\bibitem{Num01} S. Nummela \textit{et al.}, Phys. Rev. C \textbf{63}, 
                044316 (2001).

\bibitem{Cau05} E. Caurier, G. Mart\'{\i}nez-Pinedo, F. Nowacki, A. Poves, and A. P. Zuker,
                Rev. Mod. Phys. \textbf{77}, 427 (2005). 

\bibitem{Cau04} E.~Caurier, F.~Nowacki, and A.~Poves, Nucl. Phys. \textbf{A742}, 14 (2004).

\bibitem{Sor93} O. Sorlin \textit{et al.},  Phys. Rev. C \textbf{47}, 2941 (1993).

\bibitem{Sor95} O. Sorlin \textit{et al.}, Nucl. Phys. \textbf{A583}, 763 (1995).

\bibitem{Gre04} S.~Gr\'evy \textit{et al.}, Phys. Lett. \textbf{B594}, 252 (2004).


\bibitem{Rei99} P. G. Reinhard, D. J. Dean, W. Nazarewicz, J. Dobaczewski, J. A. Maruhn, 
 and M. R. Strayer,   Phys. Rev. C \textbf{60}, 014316 (1999).

\bibitem{Lal99} G. A. Lalazissis, D. Vretenar, P. Ring, M. Stoitsov, and L. Robledo,
                 Phys. Rev. C \textbf{60}, 014310 (1999).

\bibitem{Per00} S. Peru \textit{et al.}, Eur. Phys. J. A \textbf{9}, 35 (2000).

\bibitem{Ray02} R. Rodriguez-Guzm\'an, J. L. Egido, and L. M. Robledo,  
Phys. Rev. C \textbf{65}, 024304 (2002).

\bibitem{cmnp:07} E. Caurier, J. Men\'endez, F. Nowacki, and A. Poves, 
Phys. Rev. C \textbf{75}, 054317 (2007).

\bibitem{Soh02} D. Sohler \textit{et al.}, Phys. Rev. C \textbf{66}, 054302 (2002).

\bibitem{Gre05} S. Grevy \textit{et al.},  Eur. Phys. J. A \textbf{25}, 111 (2005).

\bibitem{Sar00} F. Sarazin \textit{et al.}, Phys. Rev. Lett. \textbf{84}, 5062 (2000).

\bibitem{Ibb99} R. W.~{Ibbotson}, T. Glasmacher, P. F. Mantica, and H. Scheit,  
            Phys. Rev. C \textbf{59}, 642 (1999).

\bibitem{Dom03} Zs. Dombr\'{a}di \textit{et al.}, Nucl. Phys. \textbf{A727}, 195 (2003).

\bibitem{Stu06} A. E. Stuchbery \textit{et al.}, Phys. Rev. C \textbf{74}, 054307 (2006).

\bibitem{Fri05} J.~Fridmann \textit{et al.}, Nature \textbf{435}, 922 (2005).

\bibitem{Gad05} A.~Gade \textit{et al.}, Phys. Rev. C \textbf{71}, 051301(R) (2005).

\bibitem{Fri06} J.~Fridmann \textit{et al.}, Phys. Rev. C \textbf{74}, 034313 (2006).

\bibitem{Jur06} B.~Jurado \textit{et al.}, Phys. Lett. \textbf{B649}, 43 (2007).

\bibitem{Gau06} L.~Gaudefroy \textit{et al.}, Phys. Rev. Lett. \textbf{97}, 092501 (2006).

\bibitem{Gad06} A.~Gade \textit{et al.}, Phys. Rev. C \textbf{74}, 034322 (2006).

\bibitem{Cam06} C. M.~Campbell \textit{et al.}, Phys. Rev. Lett. \textbf{97}, 112501 (2006).

\bibitem{Bas07} B. Bastin \textit{et al.}, Phys. Rev. Lett. \textbf{99}, 022503 (2007).

\bibitem{Cam07} C. M.~Campbell \textit{et al.}, Phys. Lett. \textbf{B652}, 169 (2007). 

\bibitem{nndc}  National Nuclear Data Center, Brookhaven National Lab. 2008.
                 \url{http://www.nndc.bnl.gov}.   

\bibitem{Bau98} P. Baumann  \textit{et al.}, Phys. Rev. C \textbf{58}, 1970 (1998).

\bibitem{End88} P. M. Endt and R. B. Firestone, Nucl. Phys. \textbf{A633}, 1 (1988). 

\bibitem{Uoz:94} Y. Uozumi \textit{et al.}, Phys. Rev. C \textbf{50}, 263 (1994).

\bibitem{Gel:07} M. Gelin \textit{et al.}, in  {\it Proceedings of the INPC2007 Conference} (Tokyo),
                 edited by S. Nagamiya, T. Motobayashi, M. Oka, R. S. Hayano and T. Nagae
                 (Elsevier, Amsterdam, 2008), Vol. II, p.368.  

\bibitem{Kra01} G. J. Kramer \textit{et al.}, Nucl. Phys. \textbf{A679}, 267 (2001).

\bibitem{Bro88}	B. A.~Brown and B. H.~Wildenthal, Ann. Rev. Nucl. Part. Sci. 
                \textbf{38}, 29 (1988).

\bibitem{Gui84} D. Guillemaud-Mueller \textit{et al.}, Nucl. Phys. \textbf{A426}, 37 (1984).

\bibitem{Mot95} T. Motobayashi \textit{et al.}, Phys. Lett. \textbf{B346}, 9 (1995).

\bibitem{Pov87} A. Poves and J. Retamosa,  Phys. Lett. \textbf{B184}, 311 (1987).

\bibitem{Yon01} K. Yoneda \textit{et al.}, Phys.  Lett. \textbf{B499}, 233 (2001).

\bibitem{Iwa01} H. Iwasaki \textit{et al.}, Phys.  Lett. \textbf{B522}, 227 (2001).

\bibitem{Chu05} J. A. Church \textit{et al.}, Phys. Rev. C \textbf{72}, 054320 (2005).

\bibitem{Ele06} Z. Elekes \textit{et al.}, Phys. Rev. C \textbf{73}, 044314 (2006).

\bibitem{Gade07} A. Gade \textit{et al.}, Phys. Rev. Lett. \textbf{99}, 072502 (2007).

\bibitem{bh08} S. Bhattacharyya  \textit{et al.}, Phys. Rev. Lett. \textbf{101}, 032501 (2008).


\end{thebibliography}
\end{document}